\documentclass[twocolumn,showpacs,preprintnumbers,amsmath,amssymb]{revtex4}

\usepackage{graphicx}
\usepackage{dcolumn}
\usepackage{bm}
\begin{document}
\title{Nucleation and growth in one dimension,\\ part II: Application to DNA replication kinetics}
\author{Suckjoon Jun}
\altaffiliation{Present address:  FOM Institute for Atomic and Molecular Physics (AMOLF), P.O.Box 41883, 1009 DB Amsterdam, The Netherlands}
\author{John Bechhoefer} 
\email[email: ]{johnb@sfu.ca}
\affiliation{Department of Physics, Simon Fraser University, Burnaby, B.C., V5A 1S6, Canada}%
\date{\today}

\begin{abstract}
Inspired by recent experiments on DNA replication, we apply a one-dimensional nucleation-and-growth model to DNA-replication kinetics, focusing on how to extract the time-dependent nucleation rate $I(t)$ and growth speed $v$ from data.  We discuss generic experimental problems, namely spatial inhomogeneity, measurement noise, and finite-size effects.  After evaluating how each of these affects the measurements of $I(t)$ and $v$, we give guidelines for the design of experiments.  These ideas are then discussed in the context of the DNA-replication experiments.

\end{abstract}

\maketitle

\section{Introduction}

Since its development in the late 1930s, the phenomenological model of nucleation and growth of Kolmogorov, Johnson-Mehl, and Avrami  (KJMA) has been widely applied to the analysis of kinetics of first-order phase transformations, mostly in two and three spatial dimensions~\cite{Kolmogorov, Johnson-Mehl, Avrami}.  The model has several exact results given the following basic assumptions: (1) The system is infinitely large and untransformed at time $t$=0;  (2) nucleations occur stochastically, homogeneously, and independently one from one another; (3) the transformed domains grow outward uniformly, keeping their shape; and (4) growing domains that impinge coalesce.  

Although the KJMA model is conceptually simple, experiments often have complicating factors that make the contact between theory and experiment delicate and lead to deviations from the basic model.  For example, a principal result of the KJMA model is that the fraction $f(t)$ of the transformed volume at time $t$ is
\begin{equation}
\label{eq:f}
f(t) = 1 - e^{-A t^a},
\end{equation}
where $A$ and $a$ are constants: $A$ depends upon the growth velocity $v$, the nucleation rate $I$, and the spatial dimension $D$, while $a$ is determined by $I$ and $D$.  In the literature, $a$ is called the Avrami exponent.  ``Avrami plots" of $-\ln [\ln(1-f)]$ vs. $\ln t$ should thus be straight lines of slope $a$~\cite{endnote1}.  Unfortunately, Eq.~\ref{eq:f} often does not fit data well because the experimental conditions do not satisfy the assumptions of the KJMA theory~\cite{Cahn, vanSiclen, Tomellini1997}.  For example, nucleation can be inhomogeneous or correlated~\cite{Fanfoni2003, JunCellCycle}; real systems are finite; and there is always measurement noise.

In two- or three-dimensional systems, where only limited theoretical results such as Eq.~\ref{eq:f} are available, it can be difficult to pinpoint the origins of discrepancies between experimental data and the predictions of the KJMA model.  In one-dimensional systems, however, several scientists have shown since the 1980s that one can push the analysis much further than for the original version of the KJMA model~\cite{Sekimoto, Ben-Naim1996, PaperI}.  

In this paper, we shall show that a detailed theoretical understanding of the KJMA model in 1D lets us compare theory and experiment more directly.  In other words, we can extract the kinetic parameters from data under less-than-ideal experimental circumstances.  Our discussion will be set in the context of recent DNA-replication experiments that have drawn attention from both the physics and biology communities~\cite{Herrick2000, Lucas2000, Herrick2002}.

\section{Application of the 1D-KJMA model to experimental systems}
Although there are many analytical results for the 1D-KJMA model, only a very few 1D systems that are well-described this model have been identified (e.g.~\cite{Derrida}), and very little detailed analysis has been done on those systems.  Recently, however, Herrick {\it et al.} have identified a formal analogy between the 1D-KJMA model and DNA replication processes~\cite{Herrick2002}.  Equally important, they have developed experimental methods that can yield large quantities of data, allowing the extraction of detailed statistical quantities.  Since the DNA work provides a model system for testing the general experimental problems discussed above, and also in order to fix the language, we begin by reviewing the mapping between DNA replication and the KJMA model.

\subsubsection{Mapping DNA replication onto the KJMA model}

Although the organization of the genome for DNA replication varies considerably from species to species, the duplication of most eukaryotic genomes shares a number of common features~\cite{DePamphilis}:
\begin{enumerate}
\item DNA replication starts at a large number of sites known as ``origins of replication."  The DNA domain replicated from each origin is referred to, informally, as an ``eye" or a ``replication bubble" because of its appearance in electron microscopy.
\item The position of each potential origin that is ``competent" to initiate DNA replication is determined before the beginning of the synthesis part of the cell cycle (``S phase"), when several proteins, including the origin recognition complex (ORC) bind to DNA, forming a pre-replication complex (pre-RC).
\item During S phase, a particular potential origin may or may not be activated.  Each origin is activated not more than once during the cell-division cycle.
\item DNA synthesis propagates at replication forks bidirectionally, with propagation speed or fork velocity $v$, from each activated origin.  Experimentally, $v$ is approximately constant throughout S phase.
\item DNA synthesis stops when two newly replicated regions of DNA meet.
\end{enumerate}

\begin{figure}[!t]
\centering
\includegraphics[width=3.4in]{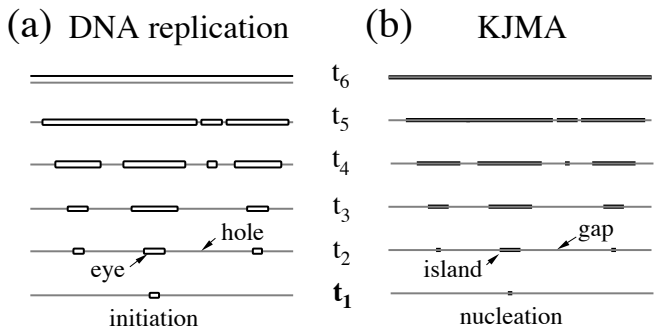}
\caption{Mapping DNA replication onto the one-dimensional KJMA model.}
\label{fig:mapping}
\end{figure}
From Fig.~\ref{fig:mapping}, it is apparent that processes 3--5 have a formal analogy with nucleation and growth in one dimension.  We identify (1) nucleation of islands as activation (initiation) of replication origins; (2) growth of the eyes as growth of the islands; and (3) coalescence of two expanding eyes as the merging of growing islands.  Of course, while DNA is topologically one dimensional, it is embodied in a three-dimensional space.

In an ideal world, one could monitor the replication process continuously and compile domain statistics  in real time.  In the real world, the three billion DNA basepairs (bps) of a typical higher eukaryote, which replicate in as many as $\sim$$10^5$ sites simultaneously, are packed in a cell nucleus of radius $\sim$1 $\mu m$, making a direct, real-time monitoring impossible~\cite{Alberts}.  Recently, experiments have used two-color fluorescent labeling of DNA bases to study replication kinetics indirectly~\cite{Herrick2000}.  One begins (in a test tube) by labeling the bases used in replicating the DNA with, say, a red dye.  At some time during the replication process (e.g. $t_1$ in Fig.~\ref{fig:mapping}), one floods the test tube with green-labeled bases and allows the replication cycle to go to completion.  One then stretches the DNA onto a glass slide (``molecular combing"~\cite{Bensimon1994}), a process that unfortunately also breaks the DNA strands into finite segments.  Under a microscope, regions that replicated before adding the dye are red, while those labeled afterwards are predominantly green.  The alternating red-and-green regions correspond to eyes and holes in  Fig.~\ref{fig:mapping}, forming a kind of snapshot of the replication state of the DNA fragment at the time the second dye was added.  Each time point in Fig.~\ref{fig:mapping} would thus correspond to a separate experiment.

Using the formal analogy between DNA replication and 1D nucleation-growth model, we can extract the kinetic parameters $I(t)$ and $v$ from data~\cite{Herrick2002}.  For the ideal case, the procedure is straightforward.  For real-world data, on the other hand, one has to be cautious because of the generic problems explained above.  We have already mentioned that the molecular combing process chops the DNA into finite-size segments, which effectively truncates the full statistics~\cite{Herrick2000}.  Another problem in the experimental protocols is that an in-vitro replication experiment usually has many different nuclei in the test tube.  These nuclei start replication at different, unknown times and locations along the genome~\cite{Herrick2000, Lucas2000}.  The asynchrony leads to sample heterogeneity and creates a starting-time distribution for the DNA replication~\cite{Herrick2002}.  Finally, the finite resolution of the microscope used to measure domain sizes may affect the statistics.

Below, we shall examine each of these complicating factors, present empirical criteria for their significance, and then discuss the implications of these criteria for the design of experiments.

To set the stage, we begin with the problem of extracting experimental parameters from ideal data.

\subsubsection{Ideal case}
From the theoretician's point of view, a system can be said to be ideal when it satisfies all underlying assumptions of the theory.  In the context of DNA replication and the KJMA model, this means that the DNA molecule is infinitely long and that the initiation rate $I$ of replication is homogeneous and uncorrelated.  Also, statistics should be directly obtainable at any time point $t$ at arbitrarily fine resolution.  Because the growth velocity of replicated DNA domains has been measured to be approximately constant, we shall limit our analysis to this special case.  One can then apply the KJMA model to a single experimental realization to extract kinetic parameters such as $I(t)$ and $v$.

In order to do this, we note that the simulation in our previous paper~\cite{PaperI} (hereafter, Paper I) is in practice such a case (system size = $10^7$, $v = 0.5$, $dt = 0.1$, $I(t) = I \cdot t$, where $I = 10^{-5}$).  Using the theoretical results obtained in Paper I, we can find an expression to invert $I(t)$ from data.  For example, the domain density $n(t)$ and the island fraction $f(t)$ at time $t$, given a time-dependent nucleation rate $I(t)$ are~\cite{PaperI}
\begin{eqnarray}
\label{eq:nS}
n(t) &=& g(t) e^{-2v \int_0^t{g(t')}dt'}\nonumber\\
f(t) &=& 1-S(t)\\ 
	&=& 1-e^{-2v \int_0^t{g(t')dt'}}.\nonumber
\end{eqnarray}

\noindent In Eq.~\ref{eq:nS}, $g(t) = \int_0^t{I(t') dt'}$, and $S(t)$ is the hole fraction.  Note that $n(t)^{-1}$ is equal to the average island-to-island distance $\bar{\ell}_{i2i}(t)$ at time $t$.  On the other hand, the average hole size $\bar{\ell}_h(t)$ is $S(t)/n(t) = g(t)^{-1}$.  Since all three domains (island, hole, and island-to-island) have equal densities $n(t)$ in one dimension, we have the following general relationship among them, which is valid even in the presence of correlations between domain sizes:
\begin{subequations}
	\begin{eqnarray}
		\label{eq:mean-field_a}
		\bar{\ell}_{i2i}(t) &=& \bar{\ell}_i(t)+\bar{\ell}_h(t)\\
		\label{eq:mean-field_b}
		f(t) &=& \frac{\bar{\ell}_i(t)}{\bar{\ell}_i(t) + \bar{\ell}_h(t)}.
	\end{eqnarray}
\end{subequations}

\noindent In other words, there are only two independent quantities among $f(t), \bar{\ell}_i(t), \bar{\ell}_h(t), \bar{\ell}_{i2i}(t)$, and we can calculate $\bar{\ell}_i(t)$ even if we do not know the exact expression for the island distribution $\rho_i(x,t)$:
\begin{eqnarray}
\label{eq:mean_q}
\bar{\ell}_i(t) &=& \frac{1}{g(t)}\big[e^{2v\int_0^t{g(t')}dt'} - 1\big]\nonumber\\
\bar{\ell}_h(t) &=& \frac{1}{g(t)}\\
\bar{\ell}_{i2i}(t) &=& \frac{1}{g(t)}e^{2v\int_0^t{g(t')}dt'}.\nonumber
\end{eqnarray}

\begin{figure}[!t]
\centering
\includegraphics[width=3.4in]{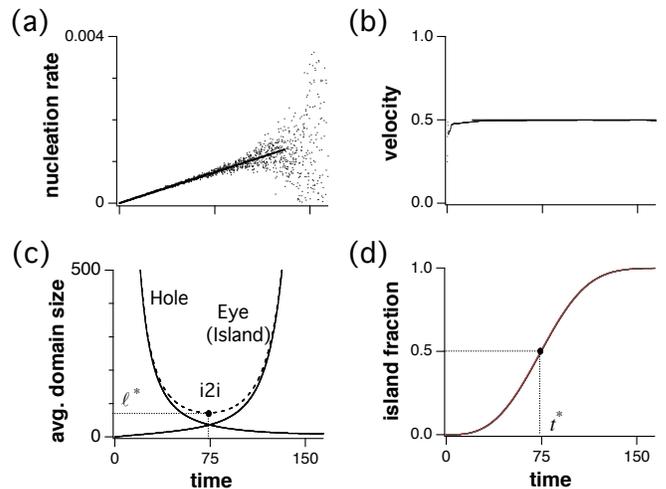}
\caption{Parameter extraction from an almost ideal data set. (a) Inferred nucleation rate vs. time; (b) Velocity vs. time; (c) Average domain sizes vs. time; (d) Island fraction vs. time; theory and extracted $f(t)$ overlap.  In (c), $\ell^*$ is the minimum average eye-to-to spacing, and sets the basic length scale.  In (d), $t^*$ is the time at which 50$\%$ of the genome has replicated.  It sets the basic time scale.}
\label{fig:basic}
\end{figure}

Note that $\bar{\ell}_i(t)$ [$\bar{\ell}_h(t)$] is a monotonically increasing (decreasing) function of time, and therefore, Eq.~\ref{eq:mean-field_a} implies that $\bar{\ell}_{i2i}(t)$ has a well-defined minimum.  We emphasize that Eqs.~\ref{eq:nS} and~\ref{eq:mean_q} set the basic time and length scales, $t^*$ and $\ell^*$, of the system.  Because the KJMA model has essentially only one scale, it is simpler than other common stochastic models in physics that lack an intrinsic scale and hence show fractal behavior (structure at all scales).  Since $f(t)$ is sigmoidal, varying from 0 to 1, we define $t^*$ to be the time required for the system to reach $f=0.5$.  On the other hand, we define $\ell^*$ to be the minimum eye-to-eye (island-to-island) distance during the course of replication [see Fig.~\ref{fig:basic}(c) and (d)].

From Eqs.~\ref{eq:nS} and~\ref{eq:mean_q}, it is straightfoward to invert the mean quantities to obtain the nucleation rate $I(t)$ and the growth velocity $v$:
\begin{eqnarray}
\label{eq:inversion1}
I(t) &=& \frac{d}{dt} \frac{1}{\bar{\ell}_h(t)}\nonumber\\
v&=& -\frac{1}{2} \frac{\ln S(t)}{ \int_0^t{{\bar{\ell}_h(t')}^{-1}dt'} }.
\end{eqnarray}

Eq.~\ref{eq:inversion1} can then be applied to an ideal set of data, i.e., one for which noise-free measurements are made on infinitely long DNA.  As Fig.~\ref{fig:basic} shows, we can recover the input parameters from simulation results in Paper I accurately: the extracted parameters  are $I = (0.99 \pm 0.04) \times 10^{-5}$ and $v=0.50 \pm 0.02$.  [The errors are the statistical errors from the curve fits in Figs.~\ref{fig:basic}(a) and (b)].  We note that the fluctuations visible for $t \agt 75$ arise from using direct numerical differentiation in Eq.~\ref{eq:inversion1}.  One could reduce the noise by appropriate data processing, using for example a smoothing spline~\cite{Recipes}.  However, because any data filtering is a delicate issue, and because direct numerical differentiation produced satisfactory results, we have decided to forego any smoothing.

We also note that there are statistical fluctuations related to the finite-size of the system: as $f(t)$ approaches 1, the number of domains $n(t)$ becomes very small; thus even small changes in $n(t)$ can cause significant fluctuations in average domain sizes.  However, the finite-size effect in this case becomes visible only when the number of new nucleations in each step, $N(t)$, is roughly 1 ($t \agt 165$ or $f \agt 0.999$).  The effect can be ignored for $N(t) \gg 1$ for the practically infinite system considered here~\cite{Cahn, Orihara}.

In the following sections, we consider the complications that arise from less-ideal experimental conditions.

\subsubsection{Asynchrony}
\begin{figure*}[!t]
\centering
\includegraphics[width=6.8in]{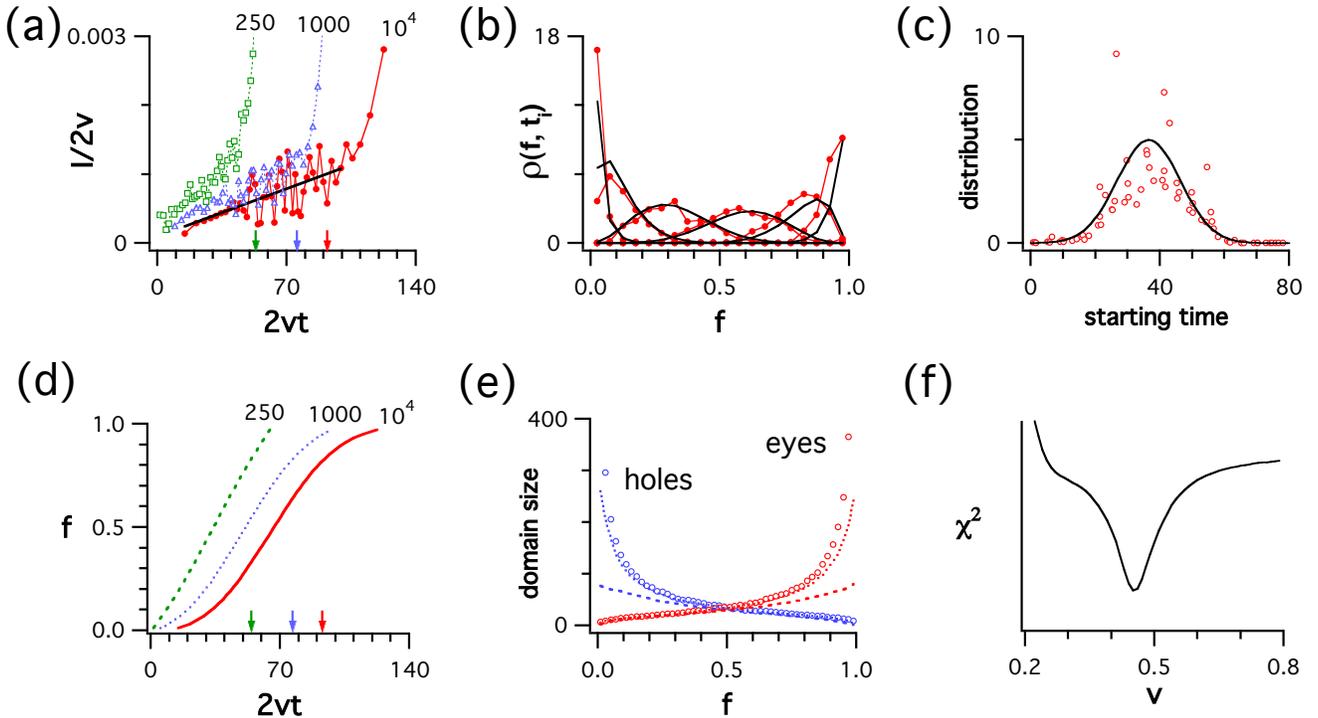}
\caption{(Color online).  Inversion results in the presence of asynchrony and finite-size effects.  (a) $I/2v$ vs. $2vt$.  The arrows indicate where $f=0.8$ in $f$ vs. $t$ curves in (d) for three different molecule sizes: $10^4$ (unchopped), 1000 and 250 (chopped).  (b) $\rho(f, t_i)$ for six time points 60, 80, 100, 120, 140, 160 (from left to right).  The circles are simulation data; the solid lines are from Eq.~\ref{eq:rhof}, using the extracted parameters in Table~\ref{tab:comparison}.  (c) Optimization results for the starting-time distribution $\phi(\tau)$.  The solid line is a Gaussian fit.  (d)  $f$ vs. $2vt$ for $\ell_c$ = 250 and $\ell_c$ = 1000.  The solid line is the unchopped case (size $10^4$).  (e) Average domain sizes vs. $f$.  The empty circles are for the unchopped case, while the dotted and dashed curves correspond to $\ell_c$ = 1000 and 250. (f) Plot of $\log{\chi^2}$ [$\rho(f, t_i)$] (arbitrary units) vs. $v$ for size $10^4$.  The complete fit results are shown in Table~\ref{tab:comparison}.  See also text.}
\label{fig:inverse1}
\end{figure*}
As we mentioned above, data often come from experiments where the DNA from many different independently replicating cells is simultaneously present in the same test tube.  The individual DNA molecules begin replicating at different unknown starting times.  In such cases, it is simpler to begin by sorting data by the replicated fraction $f$ of the measured segment~\cite{Blumenthal}.  The basic idea is that for spatially homogeneous replication (namely, nucleation and growth), all segments with a similar fraction $f$ are at roughly the same point in S phase.  Since $f(t)$ is a monotonically increasing function of $t$, we can essentially use $f$ as our initial clock, leaving the conversion to real time $t$ to a second step.

Once the data have been sorted by $f$, we extract the initiation frequency $I$ as a function of $f$.  Using Eqs.~\ref{eq:nS}-\ref{eq:mean_q}, one can straightforwardly obtain expressions analogous to Eq.~\ref{eq:inversion1}:
\begin{eqnarray}
\label{eq:inversion2}
\frac{I(f)}{2v} &=& \frac{1}{ \bar{\ell}_i+\bar{\ell}_h} \frac{d}{df} \frac{1}{\bar{\ell}_h}\nonumber\\
2v t(f) &=& \int_0^f ( \bar{\ell}_i+\bar{\ell}_h) df'.
\end{eqnarray}

\noindent In Eq.~\ref{eq:inversion2}, $\bar{\ell}_i$ and $\bar{\ell}_h$ are functions of $f$.  In other words, we have a direct inversion $I/2v$ vs. $2vt$ from data [Fig.~\ref{fig:inverse1}(a)].  Note that both $I$ and $t$ are always accompanied by the factor $2v$, which has to be determined independently (see below).
On the other hand, the fluctuations in the extracted $I/2v$ are the result of direct numerical differentiation in Eq.~\ref{eq:inversion2} discussed in the previous section.
\begin{figure*}[!t]
\centering
\includegraphics[width=6.8in]{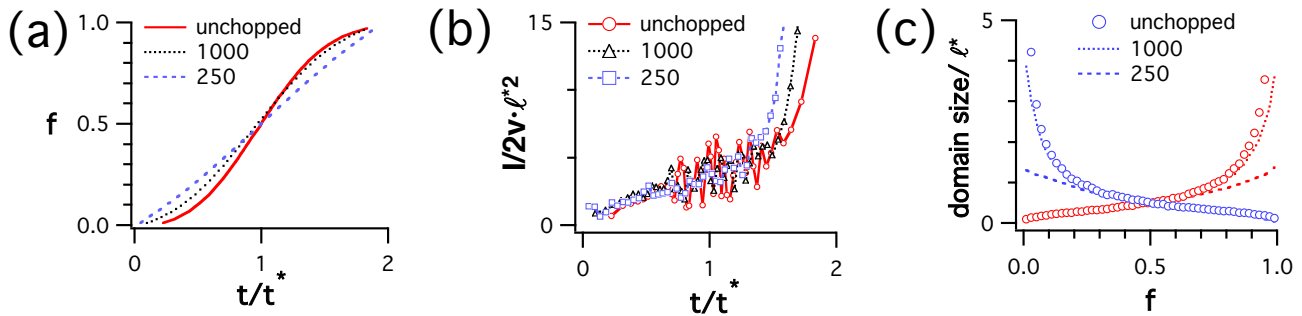}
\caption{(Color online).  Rescaled graphs for finite-size effects.}
\label{fig:rescaled_fs}
\end{figure*}

In the two-color labeling experiments, we can compile statistics into histograms of the distribution $\rho(f, t_i)$ of replicated fractions $f$ at time $t_i$ [Fig.~\ref{fig:inverse1}(b)], where $t_i$ is the timepoint where the second dye was added (Fig.~\ref{fig:mapping}).  Note that the spread in $\rho(f, t_i)$ is related to the starting-time distribution $\phi(\tau)$ via the kinetic curve $f(t)$, where $\tau$ is the laboratory time that each DNA starts replicating, and $t$ is the duration of time since the onset of replication.  Since $\phi(\tau) d\tau = \rho(f(t'), t_i) \cdot df(t')$, where $t' = t_i - \tau$, we obtain
\begin{eqnarray}
\label{eq:rhof}
\rho(f, t_i) = \phi(\tau) \times \bigg(\frac{df}{d\tau}\Big|_{t=t_i - \tau}\bigg)^{-1}.
\end{eqnarray}

For a Gaussian starting time distribution $\phi(\tau)$, one can in principle fit all $\rho(f, t_i)$'s using three fitting parameters $v$, the average starting time $\tau_0$, and the starting time width $\sigma_{\tau}$.  Unfortunately, this ``brute-force" approach did not produce satisfactory results as the basin of attraction of the minimum proved to be relatively small.

Our strategy then was first to obtain a coarse-grained $v$ vs. global $\chi^2$ plot shown in Fig.~\ref{fig:inverse1} as follows:
\begin{enumerate}
\item Guess a range of $v$ between $v_{min}$ and $v_{max}$.

\item Fix $v$ (starting from $v=v_{min}$), and trace $\rho(f, t_i)$ back in time.  For a specific value of $f$ and timepoint $t_i$, the corresponding starting time is $t_i - t(f)$ (Eq.~\ref{eq:inversion2}).  Repeat for all $\rho(f, t_i)$'s and reconstruct the starting time distribution $\phi(t)$.

\item Fit $\phi(\tau)$ obtained in step 1 to an empirical model.  (In the absence of correlations among starting times, a Gaussian distribution is a reasonable choice~\cite{endnote1.5}.  One may also know the rough form of $\phi(\tau)$ from an understanding of the origins of the asynchrony.)

\item Regenerate $\rho(f, t_i)$ using Eq.~\ref{eq:rhof} with the parameters obtained in steps 2 and 3.  Calculate $\chi^2$ for $\rho(f, t_i)$.  This is also a global fit, as the $\chi^2$ statistic is summed over data from all time points $t_i$.

\item Increase $v$ to $v$+$\Delta v$ and repeat 2--4.  If there is a well-defined minimum of the $\chi^2(v)$ (with corresponding $\tau_0$ and $\sigma_{\tau}$) [e.g., Fig.~\ref{fig:inverse1}(f)], one can find a more accurate estimate of the minimum using a standard optimization technique such as Brent's method~\cite{Recipes, IgorPro}.  Otherwise, go back to 1 and choose a different range of $v$.

\end{enumerate}
In order to test how well the optimization method described above can work in the face of asynchrony, we have repeated the simulation in Paper I with several modifications.  First, we have used 1000 molecules that started nucleations asynchronously, following a Gaussian distribution of average starting time $\tau_0=40$ and of starting time width $\sigma_{\tau}=10$~\cite{endnote:data}.  Second, the size of each individual molecule is $10^4$ instead of $10^7$.  This keeps constant the total number of ``DNA basepairs" analyzed.  

Since we used the same nucleation rate, the time to replicate to $f = 0.9$ was roughly 100 minutes, about the same as for the much larger system [see Fig.~\ref{fig:basic}(d) and Fig.~\ref{fig:inverse1}(d)].  We have chosen six timepoints ($t_i = 60, 80, 100, 120, 140, 160$) at which to collect data, and the distributions of fraction $f$ are shown in Fig.~\ref{fig:inverse1}(b).  The spread in $\rho(f, t_i)$ reflects the starting time distribution $\phi(\tau)$.

\begin{table}[!b]
\begin{tabular}{|c|c|c|}	\hline
\centering
				&	input					&	extracted						\\	\hline
~$I$~			&	~$1 \times 10^{-5}$~		&	~$(0.98 \pm 0.18) \times 10^{-5}$~	\\	\hline
$v$				&	0.5					&	0.453						\\	\hline
~starting $t$~($\tau_0 \pm \sigma_{\tau}$)~	&	~$39.6 \pm 14.1$~	&	$36.5 \pm 13.9$\\	\hline
\end{tabular}
\caption{\label{tab:comparison} Comparison between input and extracted parameters in the presence of asynchrony (starting $t$).  Note that the input $\tau_0 \pm \sigma_{\tau}$ is the Gaussian fit to a single realization of 1000 molecules, where $\tau_0 = 40$ and $\sigma_{\tau} = 10$.~\cite{endnote:data}}
\end{table}

We fit $I/2v$ vs. $2vt$ using $I(t) = a + I \cdot t$ in Fig.~\ref{fig:inverse1}(a), excluding the last few points roughly above $f=0.9$ to take into account the finite-size effect (see the following section).  We then used the fit result to obtain the growth rate $v$ by the optimization method given above.  The results are shown in Fig.~\ref{fig:inverse1} and Table~\ref{tab:comparison}.  In the plot of $\chi^2$ vs. $v$ [Fig.~\ref{fig:inverse1}(f)], we see a well-defined minimum of $\chi^2$ at $v = 0.453$, 10$\%$ below the input value 0.5.  Fig.~\ref{fig:inverse1}(b) and (c) are reconstructions of $\rho(f, t_i)$ and $\phi(\tau)$ using the parameters in Table~\ref{tab:comparison}.  The minor discrepancies in $\tau_0$ and $\sigma_{\tau}$ are acceptable, given the small number of points of $\rho(f, t_i)$ used in the optimization (20 points in each of six histograms).  Note that the finite size of sampled DNA is responsible for a larger part of the discrepancy with the original parameters than was our reconstruction algorithm.

The success of this method depends on the experimental design, as well; i.e., one has to choose the right timepoints $t_i$ in order to deduce $\phi(\tau)$ accurately [see Fig.~\ref{fig:inverse1}(b) and (c)].  The key parameter is the ratio $\alpha$ between the replication time scale $t^*$ and the starting-time width $\sigma_{\tau}$, respectively: $\alpha = t^* / \sigma_{\tau}$.  For the case considered here ($t^* \approx 75$ and $\sigma_{\tau} \approx 14$), $\alpha \approx 5.4$.

Ideally, $\alpha \gg 1$ (better synchrony with slow kinetics) so that $\rho(f, t_i)$ has a well-defined peak between $0 < f < 1$, and $\rho(f, t_i) \rightarrow 0$ as $f \rightarrow$ 0 and 1.  In this case, even a single $\rho(f, t_i)$ can be used to reconstruct $\phi(\tau)$ and extract $v$ accurately.  For example, each single histograms for all timepoints in Fig.~\ref{fig:inverse1}(b) produced results that are accurate to 15$\%$.

For $\alpha \ll 1$ (high asynchrony with fast kinetics), $\rho(f, t_i)$ is spread over $0 \leq f \leq 1$.  In this case, experimentalists should choose at least $N = \sigma_{\tau}/t^*$ timepoints to cover the whole range of $\phi(\tau)$, where well-chosen $t_i$'s spread evenly the peaks of $\rho(f, t_i)$ between 0 and 1.

\subsubsection{Finite-size effects}
As mentioned above, the DNA is broken up into relatively short segments during the molecular-combing experiments.  In order to estimate how the finite segment size affects the estimates of $I(t)$ and $v$, we have cut the simulated molecules in the previous section into smaller pieces of equal size $\ell_c$~\cite{endnote2}.  Fig.~\ref{fig:inverse1} shows results for $\ell_c$ = 1000 and 250, with original size $10^4$. As one can see, there is a clear correlation between $\ell_c$ and the statistics.  First, the smaller the segments are, the smaller the average domain sizes become as $f \rightarrow 1$.  This is as expected, since one obviously cannot observe a domain size larger than $\ell_c$.  Note that an underestimate of average eye and hole sizes, $\bar{\ell}_i$ and $\bar{\ell}_h$, leads to an overestimate of the extracted $I(t)$, as implied by Eq.~\ref{eq:inversion2}.  Second, as $\ell_c$ becomes smaller, the completion times are underestimated.  Third, the sharp increase (decrease) in average eye (hole) sizes disappears, becoming nearly flat at a characteristic fraction $f^*$, and the kinetic curve $f(t)$ significantly deviates from its sigmoidal shape, becoming nearly linear.  In fact, there is a close relationship between these last two effects.  The sharp increase in average eye size results from to the merger of smaller eyes, which dominates the late stage of replication kinetics.  Since chopping DNA eliminates the large eyes, as shown in Fig.~\ref{fig:inverse1}(e), it effectively increases the number of domains $n(t)$ per unit length in truncated segments and overestimates the replication rate. (The replication rate $df/dt = 2vn$.)

We emphasize that the first two observations above imply that $\ell_c$ affects the basic time and length scales, $t^*$ and $\ell^*$, of the (chopped) systems introduced in the previous section.  In Figs.~\ref{fig:rescaled_fs}(a)-(c), we re-plot $f(t)$, $I(t)$, and $\bar{\ell}_i$ and $\bar{\ell}_h$ using the dimensionless axes.  One can clearly see that the chopping process straightens the sigmoidal $f(t)$ and the average domain size curves.  Nevertheless, the basic shape of $I(t)$ does not change, i.e., curves corresponding to different values of $\ell_c$ collapse onto one another, and the finite-size effect only makes the up-shooting tails steeper.

\begin{figure}[!t]
\centering
\includegraphics[width=3.2in]{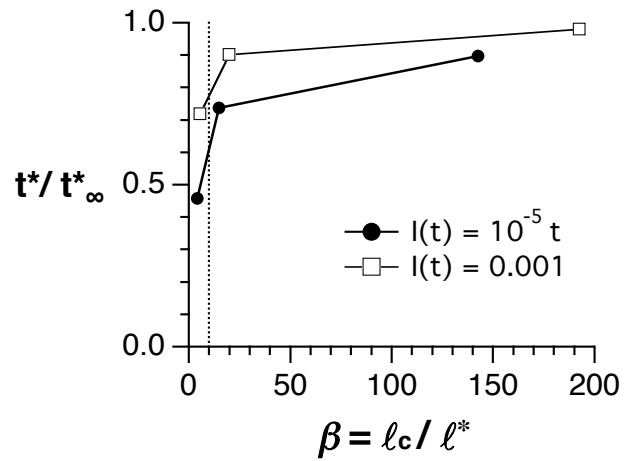}
\caption{The finite-size effects and changes in the basic time and length scales.  Shown are two different initiation rates $I(t) = 10^{-5}t$ and $I(t) = 0.001$.  The vertical line is where the average number of domains per molecule is 10.  The y-axis has been normalized relative to the initiation rate for an infinite system ($\beta \rightarrow \infty$).}
\label{fig:fs_ref}
\end{figure}

As criteria for significance of finite-size effects, we first define a new parameter $\beta = \ell_c/\ell^*$, namely, the maximum average number of domains per chopped molecule (around $f=0.5$).  Then, more careful observation of Figs.~\ref{fig:rescaled_fs}(a) and (c) suggests that there might exist a critical value $\beta^*$ (or corresponding chopping size $\ell_c^*$), where the finite-size effects severely affect the statistics.  In other words, for $\beta > \beta^*$, one can ignore the finite-size effects by excluding the last few data points close to $f=1$ (Recall that $\ell^*$ is the minimum average eye-to-eye spacing).  To see this clearly, in Fig.~\ref{fig:fs_ref}, we have plotted $t^*/t^*_{\infty}$ vs. $\beta$ for two different cases: $I(t) = 10^{-5}t$ and $I(t)=0.001$, where $t^*_{\infty}$ has been calculated using the basic kinetic curve $f(t) = 1-\exp[-2v\int_0^t{g(t')dt'}]$ (i.e., the system is infinitely large)~\cite{endnote1, PaperI}.

Indeed, changes in $t^*$ are very slow above $\beta \approx 10$, but drop sharply below this ratio.  Since $\beta$ is the average number of domains per molecule, we argue that the KJMA model can be applied to data directly when there are enough eyes in individual molecule fragments (roughly, at least 10).  On the other hand, when $\beta \alt 10$, one would require more sophisticated theoretical methods to obtain correct statistics.

\begin{figure*}[!th]
\centering
\includegraphics[width=6.3in]{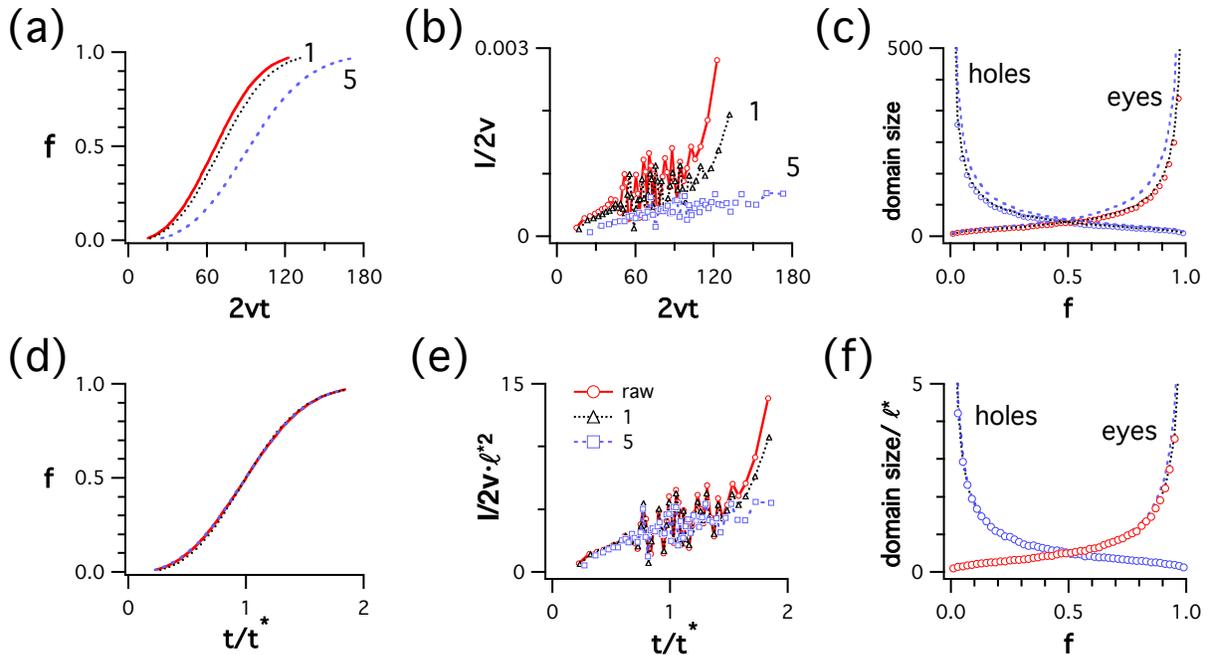}
\caption{(Color online).  The effect of coarse-graining.  (a) $f$ vs. $2vt$.  From left to right, $\Delta x^*$ = 0, 1, 5.  (b) $I/2v$ vs. $2vt$.  From top to bottom, the coarse-graining factor $\Delta x^*$ = 0 (no coarse-graining), 1 (comparable to optical resolution), and 5.  (c) Average domain sizes vs. $f$.  The empty circles are for no coarse-graining, while the dashed lines are for $\Delta x^*$ = 1 and 5 (dotted and dashed, respectively). (d)-(f) Rescaled graphs.}
\label{fig:resolution}
\end{figure*}

One subtle point is that $t^*$, unlike $\ell^*$, is not very accessible experimentally and requires data processing for accurate extraction [e.g. Fig.~\ref{fig:inverse1}(d) or Fig.~\ref{fig:resolution}(b)].

Finally, we note that the sudden up-shooting in the tails of the extracted $I(t)/2v$ vs. $2vt$ curves are yet another kind of finite-size effect related to numerical differentiation (Eq.~\ref{eq:inversion1}).  This can be simply excluded from the analysis.

\subsubsection{Finite-resolution effect}
Another generic problem is the finite resolution of measurements.  In molecular combing experiments, for example, epifluorescence microscopy is used to scan the fluorescent tracks of combed DNA on glass slides.  The spatial resolution ($\sim$1 kb) means that smaller domains will not be detectable.  Thus, two eyes separated by a hole of size $\leq$ 1 kb will be falsely assumed to be one longer eye.  We evaluate this effect by coarse-graining the statistics with experimental resolutions $\Delta x^*$, while keeping $\Delta x = v \cdot dt$ in simulation much finer.  To coarse grain by a factor $\delta = \Delta x^*/\Delta x$, we have used the raw, ``unchopped" data set in the previous finite-size-effect section:  after the simulation, we have scanned the final lists of eyes and holes, $\{i\}$ and $\{h\}$, and removed any eyes (holes) for $\delta < 1$, combining them with the two flanking holes (eyes) into a larger hole (eye) that equals the size of all three domains.

In Figs.~\ref{fig:resolution}(a)-(c), we show how the statistics change by coarse-graining only (i.e., without chopping), where the coarse-graining factors $\delta$ are 20 and 100.

The finite-resolution effect biases estimates in a way that is opposite to finite-size effects, i.e., converting eyes (holes) for $\delta < 1$ to holes (eyes) increases the average domain sizes.  As a consequence, the extracted $I(t)$ is slightly underestimated.  Nevertheless, the curves in each of $f(t)$, $I(t)$, and $\bar{\ell}_i$ and $\bar{\ell}_h$ almost perfectly collapse onto each other when the axes are rescaled using $t^*$ and $\ell^*$, confirming that, as with finite-size effects, the main consequence is a change in the basic time and length scales of the problem [Fig.~\ref{fig:resolution}(d)-(f)].

To find criteria for significance of finite-resolution effects, we recall that coarse-graining falsely eliminates eyes and holes smaller than the resolution $\Delta x^*$ only ($\delta < 1$).  For example, statistics for $f$$\approx$$0$ (small eyes) or $f$$\approx$$1$ (small holes) can be affected by coarse-graining.  For these two cases, however, one can easily avoid a problem by excluding data for $f \approx$ 0 and 1 from analysis.  

On the other hand, a more serious situation can arise when $\gamma = \ell^*/\Delta x^* \alt 1$, because a resolution comparable to the minimum eye-to-eye distance will seriously alter the mean domain sizes $\bar{\ell}_i$ and $\bar{\ell}_h$ and thus the extracted $I(t)$, as well.  Indeed, for $\gamma \gg 1$, the $\rho(f, t_i)$'s remain essentially unchanged (i.e., the optimization result for $v$ remains the same) even at $\delta = 100$ (where, $\gamma \approx 70$) (data not shown).  We conclude that $\gamma = 1$ is the relevant criterion to test the significance of finite-resolution effects.

\section{Discussion and Conclusion}
In the previous section, we have tested various generic experimental limitations via Monte Carlo simulation.  When the system is large ($10^7$ for $v=0.5$ and $I(t) = 10^{-5}t$), we have been able to extract all the input parameters accurately from a single realization of our simulation.  As the experimental (simulation) conditions become less ideal, however, one requires more sophisticated tools.  

In the presence of asynchrony, we have demonstrated that the input parameters can still be extracted to reasonable accuracy (roughly $10\%$ for $\alpha \approx 5.4$) using an optimization method.  In most DNA replication experiments, $\alpha \agt 1$.  For example, in the {\it Xenopus} egg extracts experiments of Herrick {\it et al.}~\cite{Herrick2000, Herrick2002}, $\alpha \approx 2.5$ ($t^* \approx 15$ mins and $\sigma_{\tau} \approx 6$ mins).  In this case, the method presented here can even be applied to data $\rho(f, t_i)$ for a single well-chosen timepoint $t_i$ to extract $v$.  The accuracy increases as more data are collected for different timepoints.

The significance of finite-size effects can be estimated by the criterion $\beta = \ell^*/\ell_c \approx 10$.  Fortunately, $\ell^*$ for {\it Xenopus} sperm chromatin is roughly 10 kb, while the typical size of combed molecules ranges between 100 - 500 kb, thus giving $10 \alt \beta \alt 50$.  However, the origin spacing of many higher eukaryotes, including {\it Xenopus} after the mid-blastula transition, can be as large as 100 kb.  In such cases, it is of critical importance to obtain long combed molecules ($>$ 1 Mb).

Similarly, finite-resolution effects are insignificant when $\gamma =  \ell^*/\Delta x^* > 1$.  This condition is satisfied in almost all molecular-combing experiments of DNA replication, since $\Delta x^* \approx 1$ kb while $\ell^*$ typically ranges between 10 and 100 kb ($\gamma \approx$ 10 to 100).

Among the various experimental limitations we have tested, the finite-size effects seem to be potentially the most serious problem in the molecular-combing experiments.  Fortunately, we expect the finite-size effects in the experiments and analysis of refs.~\cite{Herrick2000, Herrick2002} to be relatively insignificant because $\beta > 10$.  On the other hand, we need more sophisticated theoretical tools to correct the finite-size effects for $\beta < 10$.  We recall that the coarse-graining of molecules affects the tails in Fig.~\ref{fig:resolution}(b) opposite to the way the finite-size of molecules affects them.  We thus speculate that an intelligent way of annealing finite-sized molecules can reduce or correct the finite-size effects.  We leave a detailed evaluation of this idea for future work.

In summary, we have discussed how to apply the KJMA model to data to extract kinetic parameters under various experimental limitations, such as asynchrony, finite-size, and finite-resolution effects.  For the application to DNA-replication experiments, we have shown that finite-size effects can be ignored when the chopped molecules contain enough domains (i.e., $\beta \agt 10$).  Even when the size of molecules is smaller than the critical value $\ell_c^*$, the shape of the nucleation rate $I(t)$ is not affected when plotted using rescaled parameters.  On the other hand, finite-resolution effects are insignificant when $\gamma \gg 1$, which is the case for molecular combing experiments of DNA replication.

The theoretical understanding of these limitations given here should provide guidelines for the design of future experiments.\\

\begin{acknowledgments} 
We thank Aaron Bensimon and John Herrick for collaboration in the interpretation of their experiments on DNA replication, and we thank Tom Chou, Massimo Fanfoni, Govind Menon, Nick Rhind, and Ken Sekimoto for helpful comments and discussions on 1D nucleation-and-growth models.  This work was supported by NSERC (Canada).
\end{acknowledgments}


\begin{thebibliography}{20}
\bibitem{endnote1}Eq.~\ref{eq:f} comes from the more general expression $f(t) = 1 - \exp[-v^D \int{I(\vec{x}, t) d^Dx}]$, where the integral is performed over the so-called extended volume.  For $D=3$ and $I(\vec{x}, t) = I =$ const., one obtains $f(t) = 1-\exp{(-\frac{\pi}{3} I v^3 t^4)}$, giving $A=\frac{\pi}{3} I v^3$ and $\alpha=4$.  Note that different values of $A$ are related by rescalings of $v$ and $I$.
\bibitem{endnote1.5}{Since the only relevant parameters of $\phi(\tau)$ are its mean and standard deviation, maximum-entropy arguments also justify the choice of a Gaussian distribution~\cite{Sivia}.}
\bibitem{endnote:data}We note that the actual realization of starting times in the particular simulation of Table~\ref{tab:comparison} is $\tau_0 = 39.6$ and $\sigma_{\tau} = 14.1$.  In other words, there are always errors related to the amount of data used in analysis, but this is a separate issue from the extraction methods presented here.
\bibitem{endnote2}Experimentally, the size distribution of DNA fragments is Poissonian.  Similar finite-size effects obtained for any unimodal distribution with the same mean.
\end{thebibliography}
\end{document}